# Insertion of CdSe quantum dots in ZnSe Nanowires: MBE growth and microstructure analysis

M. den Hertog[1], M. Elouneg-Jamroz[1], E. Bellet-Amalric[2], S. Bounouar[1], C. Bougerol[1], R. André[1], Y. Genuist[1], J.P. Poizat[1], K. Kheng[2] and S. Tatarenko[1]

*Nanophysics and Semiconductors Group*
*1) Institut Néel, CNRS, BP 166, 25 rue des Martyrs, F-38042 Grenoble Cedex 9, France and*
*2) CEA, INAC, SP2M, 17 rue des Martyrs, F-38054 Grenoble Cedex 9, France*

**Abstract**

ZnSe nanowire growth has been successfully achieved on ZnSe (100) and (111)B buffer layers deposited on GaAs substrates. Cubic [100] oriented ZnSe nanowires or [0001] oriented hexagonal NWs are obtained on (100) substrates while [111] oriented cubic mixed with [0001] oriented hexagonal regions are obtained on (111)B substrates. Most of the NWs are perpendicular to the surface in the last case. CdSe quantum dots were successfully incorporated in the ZnSe NWs as demonstrated by transmission electron microscopy, energy filtered TEM and high angle annular dark field scanning TEM measurements.

ZnSe nanowires (NWs) with CdSe quantum dot (QD) insertions are promising objects for opto-electronic applications, for example due to the large exciton binding energy and strong carrier confinement. ZnSe NWs can be prepared by vapor phase growth [1], metalorganic chemical vapor deposition [2], molecular-beam epitaxy (MBE) [3,4] or thermo-chemical processes [5]. With the ability to precisely control growth parameters and to accurately monitor the growth process, MBE is an ideal tool to grow nano-structured materials.

We have recently shown that a single CdSe quantum dot (QD) embedded in a ZnSe NW is an efficient single photon source operating at temperatures as high as 220K [6]. However, when grown on an oxidized Si (100) substrate in the VLS growth mode catalyzed by gold particles, the NWs present a random distribution of orientations. Ohno *et al*. reported MBE growth of ZnSe NWs on ZnSe/GaAs templates by using Fe as catalyst but no clear epitaxial relationship between the template and the NWs was observed [7]. In this contribution we report on the epitaxial growth of ZnSe and ZnSe-CdSe-ZnSe QD embedded heterostructured NWs deposited on 2D ZnSe (100) or ZnSe (111)B buffer layer epitaxially grown on GaAs (100) or a GaAs (111)B substrates respectively.

The NWs are grown by MBE by using gold as a catalyst. Samples used in this study were all grown in a Riber 32P solid source MBE system. The source materials for the MBE system were elemental Zn, Se and Cd. GaAs (100) and (111)B commercial wafers were used as substrates. The GaAs substrates are deoxidized under As flow. Then a GaAs buffer layer is grown by MBE in a connected III-V MBE chamber. In order to avoid Ga incorporation in the NWs and to improve the expitaxial relation between NW and substrate, a thin ZnSe buffer layer (about 30 nm thick) is grown at 280°C on the GaAs epi-layer. Gold is then evaporated on the ZnSe buffer layer at room temperature in a dedicated metal deposition chamber connected to the II-VI and III-V growth chambers by UHV path. Dewetting of the gold film was done at 500-530°C for several minutes. The ZnSe NWs are grown at different temperatures between 300°C and 450°C under excess of Se flux. The beam equivalent pressure ratio Zn:Se is 1:4 and the pressure in the $10^{-7}$ Torr range. The CdSe insertions are grown during the NW epitaxial growth process by switching the Zn flux to a Cd flux for 30s.

The Cd:Se ratio is 1:3. A JEOL 3010 was used for high resolution transmission electron microscopy (HRTEM) in combination with energy filtered TEM (EFTEM) and a probe corrected FEI Titan was used for high angle annular dark field scanning TEM (HAADF STEM).

The gold dewetting process on ZnSe has been studied previously in detail: nanotrenches with gold particles inside [8, 9] are formed at 530°C (Fig. 1a). Gold particles with a diameter of a few nm are localised at the extremity of the nanotrenches. However when dewetting at 500°C, gold particles are located randomly on the terraces, as observed by Scanning Electron Microscopy (SEM) (Fig. 1b). The density of the gold nanoparticles can be adjusted by the Au deposition time between $10^9/cm^2$ and $50 \times 10^{10}/cm^2$. In the present work we have used a dewetting process at 500°C in order to avoid the growth of the NWs in different directions as it was observed when nanotrenches are present [9].

When a GaAs buffer is grown prior to the ZnSe buffer deposition and NW growth, cubic [100] oriented ZnSe NWs are obtained, while [0001] oriented hexagonal ZnSe NWs are obtained for the growth on a ZnSe buffer layer deposited directly on the GaAs (100) substrate that was flash-deoxidized at 580°C. The role of the GaAs buffer layer is to reduce the surface roughness, ensuring planar growth. A manifestation of planar growth is the presence of clear reflection high energy electron diffraction (RHEED) oscillations during the ZnSe buffer layer growth. SEM and HRTEM images (Fig 2) show ZnSe NWs grown on (100) ZnSe template. They are oriented along the [100] axis and present a cubic structure (Fig. 2c). The epitaxial relationship between the buffer and a ZnSe NW is clearly visible in Fig 2c. NWs grown on ZnSe (111)B templates are perpendicular to the surface and are composed of mixed [111] oriented cubic and [0001] oriented hexagonal regions (Fig. 3). Independent of the substrate, the NWs have diameters around 10 nm, close to the bulk CdSe exciton Bohr diameter (11nm).

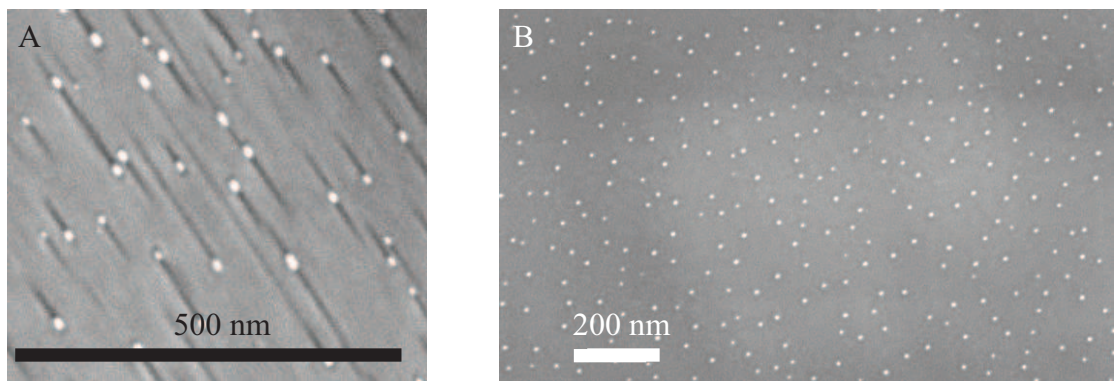

Fig 1 SEM images of 30nm ZnSe buffer layer with a 0.1 nm thick Au layer heated (A) at 530°C and (B) at 500°C

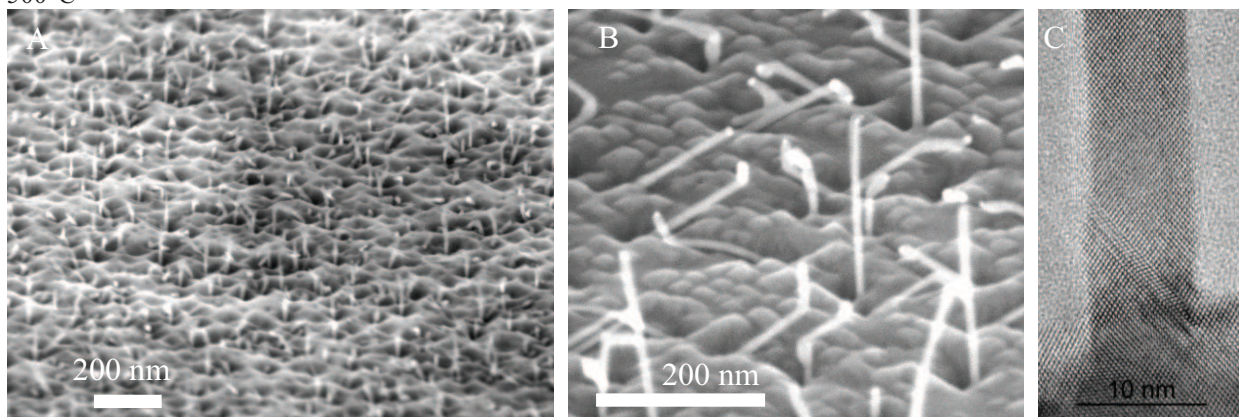

Fig. 2. SEM images (A,B) and TEM image (C) of cubic ZnSe NWs grown on a ZnSe/GaAs (100) buffer layers.

The ZnSe NWs with CdSe QD insertion were studied in detail by HRTEM, HAADF STEM, EFTEM and Energy Dispersive X-ray spectroscopy (EDX) measurements, allowing us to characterize the crystalline structure, the chemical composition and the size of the CdSe QDs. HRTEM images were treated by the Geometrical Phase Analysis (GPA) method in order to get the interplanar spacing variations along the wire axis [9]. Since the (111) interplanar distance is larger in CdSe than in ZnSe the QD position is marked by a larger lattice spacing equal to the expected lattice spacing in $CdSe_{111}$. GPA shows that the QD size in the considered NW sample is around 2 nm. EFTEM images demonstrate the presence of Cd in the QD region (Fig. 4b) and the absence of Zn in the same region (Zn map in Fig. 4c), in a NW grown at 400°C. In Fig. 4 a thin shell of 2-3 nm of ZnSe can be observed around the QD. The length of the QD is around 15 nm. Interestingly, for the same deposition time and fluxes of Cd and Se, the CdSe inclusions are as small as 1.5 nm when growth occurs at 450°C [9]. It will be possible to take advantage of this property to control the size of the QDs. We speculate that the QD length difference is due to an increased re-evaporation of Cd (or lower sticking coefficient) in the 400-450°C range (in comparison we can grow pure CdSe NWs only at or below 350°). The CdSe inclusions systematically present a cubic zinc-blende arrangement with [111] as the growth axis when inserted in hexagonal ZnSe NWs [9] and [100] as the growth axis when inserted in cubic [100] ZnSe NWs (Fig 5). In the case of hexagonal ZnSe NWs, due to the presence of 4 equivalent [111] directions, the modification in the stacking sequence at the inclusion level can explain the change in the orientation of some NWs as observed on SEM (Fig. 2b) and TEM images (Fig. 4). Interestingly bulk ZnSe has the cubic crystal structure while bulk CdSe has a hexagonal structure.

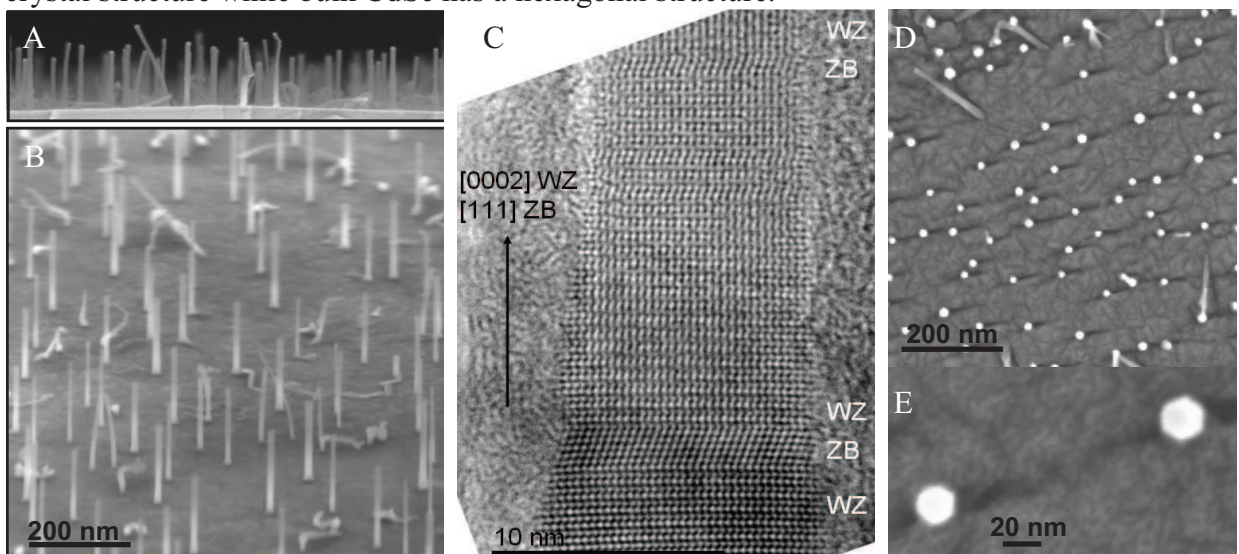

Fig. 3. SEM images (A,B) and TEM image (C) of a ZnSe NW containing hexagonal (WZ) and cubic (ZB) regions grown on a ZnSe (111) B buffer layer. (D,E) SEM top view.

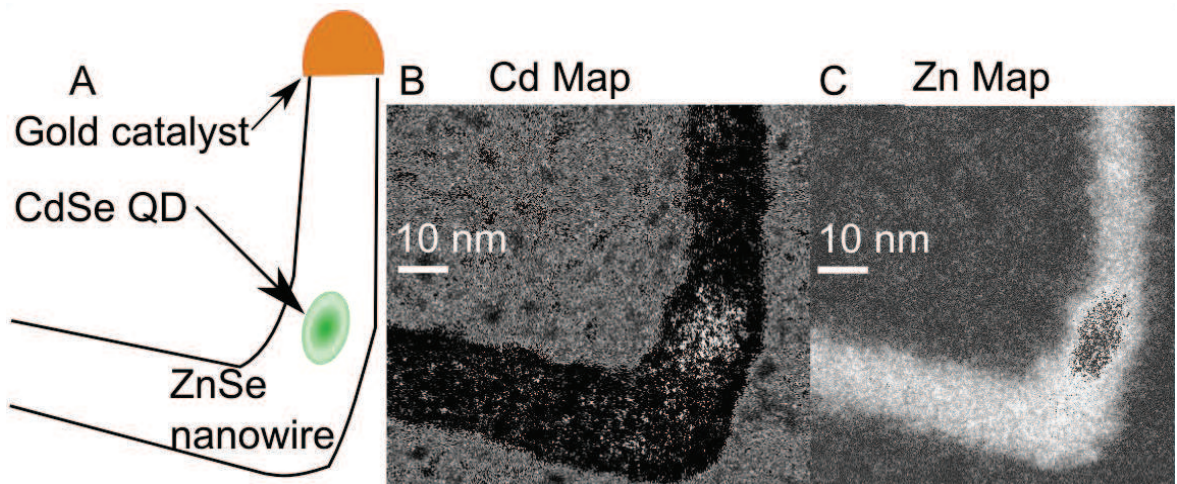

Fig 4. Composition maps of Cd and Zn obtained on a ZnSe NW with CdSe QD using EFTEM. Bright contrast indicates the presence of Cd and Zn respectively. (A) Schematic of the NW outline with CdSe QD. (B) Cd composition map (using the Cd M edge at 404 eV). (C) Zn composition map (using the Zn L edge at 1020 eV). A thin shell (2 nm) of ZnSe is visible around the QD.

Fig. 5a shows a HAADF STEM image of a cubic ZnSe NW oriented in the [100] direction with a CdSe QD with the same structure. An intensity profile (Fig. 5b) along the wire allows estimating the QD size to 3 nm, and indicates an interdiffusion region of ZnCdSe at the sides of the QD. Photoluminescence (PL) measurements have been performed on NWs detached from the growth substrate by contact with a Si substrate. The broken NWs are then lying on this Si substrate enabling the PL experiments. An example of a PL spectrum obtained on at least two distinct NWs from the sample analyzed in Fig. 5a is shown in Fig. 5c. The observed peaks are exciton-biexciton PL of 2 QDs in two separate NWs, that are excited within the excitation laser spot (typical diameter of 1 μm) at the same time. These PL results exhibit the typical signature of CdSe/ZnSe QDs [10]. This shows therefore that a CdSe QD has indeed been inserted in the ZnSe NWs, and that the number of defects inducing non-radiative recombinations is sufficiently low to allow for efficient light emission.

The CdSe QD insertion was also visualized by HAADF STEM in NWs grown on GaAs(111)B substrates (not shown). However no PL was observed from these samples, which is attributed to the high defect density in NWs grown on (111) substrates (see for example Fig. 3c). In future work the crystalline quality of NWs grown on (111) substrates should be improved.

Combining PL and TEM measurements, the PL energy depends on the length of the QDs. PL peaks in the 500, 550 and 600 nm regions were observed for QDs with a length of 3, 6 and 15 nm respectively, indicating a correlation between QD size and emission energy consistent with an increase in the exciton confinement energy in QDs of decreasing size.

    To summarize: Depending on the presence of a GaAs buffer layer deposited prior to ZnSe buffer layer and NW growth, [100] oriented cubic ZnSe NWs or [0001] oriented hexagonal ZnSe NWs are obtained for the growth on GaAs (100). On ZnSe and GaAs buffer layers deposited on GaAs (111)B NWs with mixed [111] oriented cubic and [0001] oriented hexagonal regions are obtained. CdSe insertions in the ZnSe NWs were successfully achieved and exhibited exciton-biexciton photoemission in the 500 - 600 nm region. The size of the CdSe insertions can be controlled by change of the growth temperature for a fixed CdSe growth time. The structure of the CdSe QD is cubic.

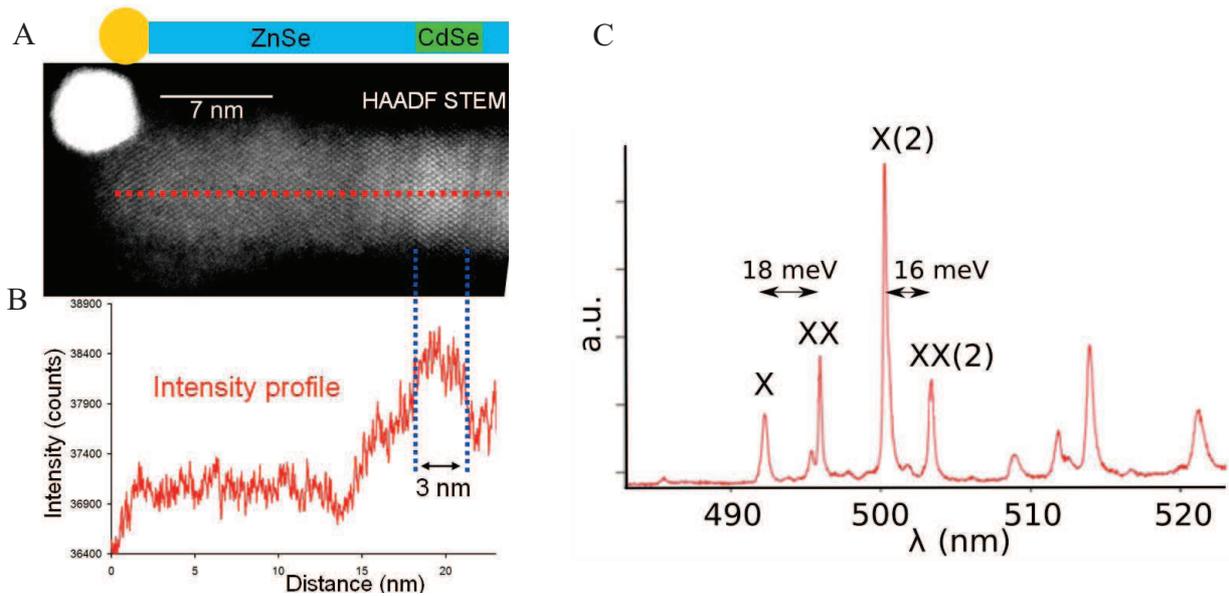

Fig. 5. (A) HAADF STEM image of a cubic NW oriented in the [100] direction with a CdSe QD (brighter contrast) with the same crystalline phase and (B) Intensity profile along the NW. (C) PL spectrum of at least two NWs from the same sample at an excitation wavelength of 405 nm at 4K. The two peaks labeled X and XX correspond to an individual QD in a NW, and the peaks labeled X(2) and XX(2) correspond to another QD in another NW.

Acknowledgements: This work is supported by French National Agency (ANR) through Nanoscience and Nanotechnology Program (Project BONAFO n°ANR-08-NANO-031-01). M E-J acknowledges financial support of the "Nanosciences aux Limites de la Nanoélectronique" Foundation.